\documentclass[runningheads]{llncs}
\usepackage{subfigure}
\usepackage{amsmath,amssymb}
\usepackage{graphicx} \graphicspath{ {Images/} }
\usepackage{amsmath}
\usepackage{booktabs}
\usepackage{microtype}

\usepackage{xcolor}
\usepackage{float}
\usepackage[colorlinks = true,
            linkcolor = blue,
            urlcolor  = blue,
            citecolor = blue,
            anchorcolor = blue]{hyperref}
\begin{document}
\title{UNetFormer: A Unified Vision Transformer Model and Pre-Training Framework for 3D Medical Image Segmentation}
\titlerunning{UNetFormer: Vision Transformer for Medical Image Segmentation}

\author{Ali Hatamizadeh \and Ziyue Xu \and Dong Yang \and Wenqi Li \and \\ Holger Roth \and Daguang Xu}
\authorrunning{Hatamizadeh et al.}
\institute{NVIDIA, Santa Clara, CA, USA \\
\email{\tt\small \{ahatamizadeh, ziyuex, dongy, wenqil, hroth, daguangx\}@nvidia.com}}

\maketitle              
\begin{abstract}
Vision Transformers (ViT)s have recently become popular due to their outstanding modeling capabilities, in particular for capturing long-range information, and scalability to dataset and model sizes which has led to state-of-the-art performance in various computer vision and medical image analysis tasks. In this work, we introduce a unified framework consisting of two architectures, dubbed UNetFormer, with a 3D Swin Transformer-based encoder and Convolutional Neural Network (CNN) and transformer-based decoders. In the proposed model, the encoder is linked to the decoder via skip connections at five different resolutions with deep supervision. The design of proposed architecture allows for meeting a wide range of trade-off requirements between accuracy and computational cost. In addition, we present a methodology for self-supervised pre-training of the encoder backbone via learning to predict randomly masked volumetric tokens using contextual information of visible tokens. We pre-train our framework on a cohort of $5050$ CT images, gathered from publicly available CT datasets, and present a systematic investigation of various components such as masking ratio and patch size that affect the representation learning capability and performance of downstream tasks. We validate the effectiveness of our pre-training approach by fine-tuning and testing our model on liver and liver tumor segmentation task using the Medical Segmentation Decathlon (MSD) dataset and achieve state-of-the-art performance in terms of various segmentation metrics. To demonstrate its generalizability,  we train and test the model on BraTS 21 dataset for brain tumor segmentation using MRI images and outperform other methods in terms of Dice score. \\
Code: \href{https://github.com/Project-MONAI/research-contributions}{https://github.com/Project-MONAI/research-contributions}
\end{abstract}

\section{Introduction}
Recently, Vision Transformers (ViT)s~\cite{dosovitskiy2020image} have achieved state-of-the-art performance in various computer vision~\cite{liu2021swin,cheng2021per} and medical image analysis tasks~\cite{hatamizadeh2022unetr,shamshad2022transformers}. In comparison to Convolutional Neural Networks (CNN)s, ViTs learn more uniform representations across their architecture with better capability in modeling long-range dependencies~\cite{raghu2021vision}. Swin Transformers~\cite{liu2021swin} were proposed to address shortcomings of ViT-based models, such as fixed token resolution and lack of inductive bias, by introducing a hierarchical architecture with patch merging layers and relative position embedding. In medical image analysis, a number of efforts~\cite{chen2021transunet,wang2021transbts} proposed to use ViTs as a standalone layer in a CNN-based U-shaped architecture. UNETR~\cite{hatamizadeh2022unetr} demonstrated that using a ViT-based encoder, which is connected to a CNN-based decoder, has a better performance by directly utilizing 3D tokens. Despite such achievements, the fixed size of visual tokens in UNETR requires additional convolutional layers to project representations at each stage to a given resolution, hence increasing the number of parameters and computational cost. nnFormer~\cite{zhou2021nnformer} proposed to use a hierarchical encoder-decoder architecture with symmetric design that consists of interleaved CNN and Swin Transformer-based blocks. However, such a model may become computationally expensive, as shown in our analysis, and perform sub-optimally in certain tasks, such as masked image modeling, due to the limited receptive field of CNN layers.

Furthermore, following the success of recent advances in masked language modeling and auto-regressive pre-training in NLP, a number of recent efforts~\cite{bao2021beit,xie2021simmim,he2021masked} have demonstrated the feasibility of self-supervised BERT-style~\cite{devlin2018bert} visual representation learning with outstanding performance in downstream dense prediction tasks such as image segmentation. However, similar techniques for self-supervised pre-training of transformer-based models have not been explored for medical image analysis. Since data annotation in the medical domain is prohibitively time-consuming and requires knowledge expertise, such pre-training schemes may become very useful as they could bolster the performance of downstream tasks (\textit{e.g.} segmentation) and allow for training more data-efficient models.   

In this work, we propose a unified framework consisting of hybrid and transformer-based architectures, referred to as UNetFormer and UNetFormer+, which utilize a 3D Swin Transformer as the encoder and CNN-based and Swin Transformer-based decoders respectively. We also propose a self-supervised pre-training scheme in which the encoder reconstructs randomly masked tokens by leveraging the visible patches. We have validated the effectiveness of our proposed framework by pre-training on a cohort of $5050$ CT images and fine-tuning on Medical Segmentation Decathlon (MSD) dataset~\cite{antonelli2021medical} for the task of liver and tumor segmentation and achieved state-of-the-art performance. In addition, we have investigated various components of the pre-training scheme for optimal representation learning. Without pre-training, we have also shown the effectiveness of the proposed architecture on the task of brain tumor segmentation of BraTS 21 dataset~\cite{baid2021rsna} and outperformed competing approaches by a large margin in terms of various performance metrics.

\section{Method}
\begin{figure}[t]
\centering
\includegraphics[width=0.9\textwidth]{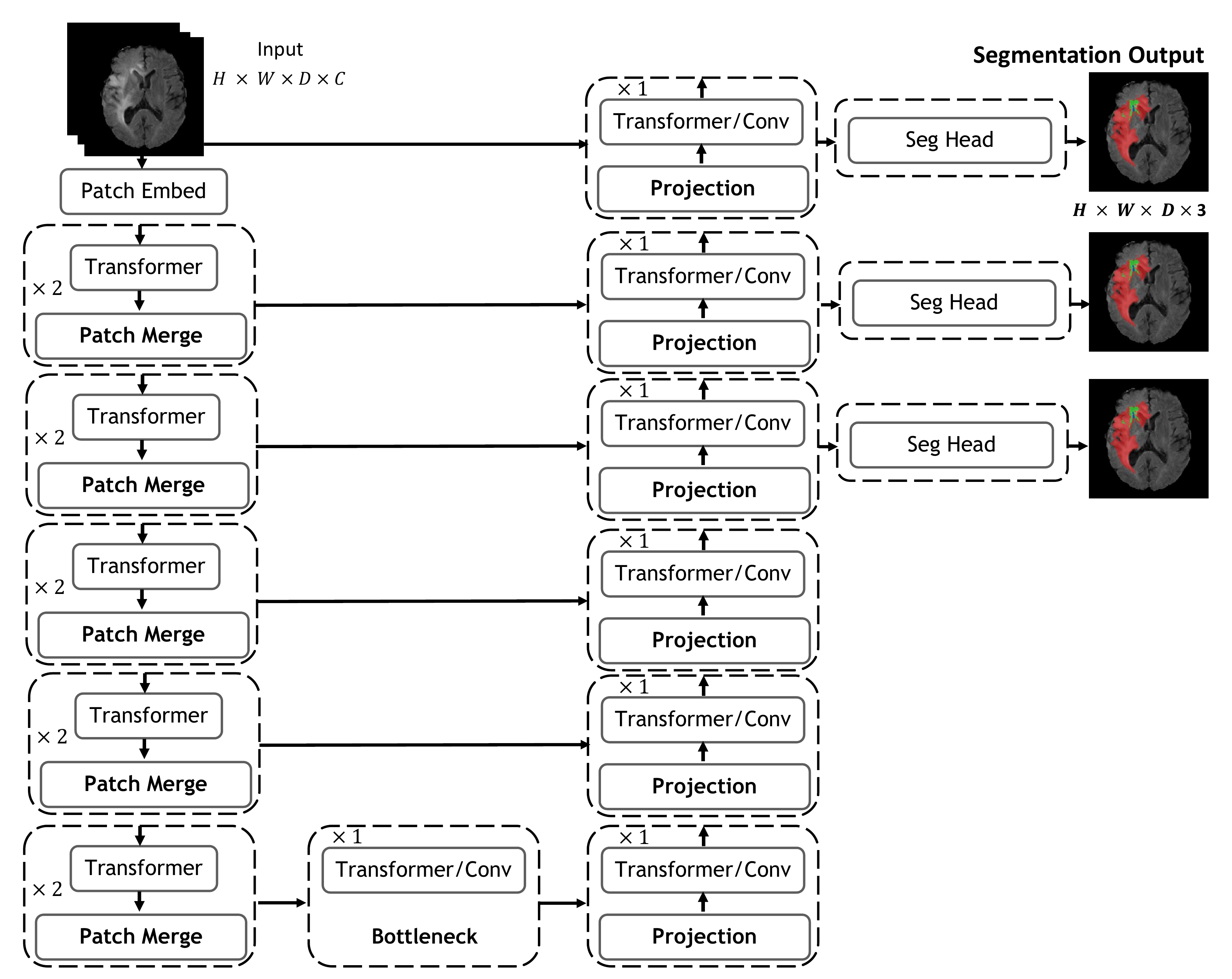}
  \caption{Overview of the UNetFormer architecture. Components of CNN and transformer-based decoders for UNetFormer and UNetFormer+ respectively are shown side by side. Seg Head denotes the final segmentation layer.}
  \label{fig:pipeline}
\end{figure}
\subsection{Model Architecture}
The architecture of UNetFormer and UNetFormer+ with CNN-based and transformer-based decoders respectively is illustrated in Fig.~\ref{fig:pipeline}. The proposed framework utilizes a 3D Swin Transformer encoder which extracts feature representations at different stages. Given a 3D input image $\mathbf{x} \in \mathbb{R}^{H\times{W}\times{D}\times{1}}$, without loss of generality, we assume that each 3D token has a size of $2\times 2\times 2\times 1$. Hence, a patch partition layer creates $\frac{H}{2}\times\frac{W}{2}\times\frac{D}{2}$ tokens with $8$ dimensional feature space that are projected into a $C$ dimensional embedding space. Each transformer block consists of two consecutive layers which compute self-attention in a local manner. Specifically, assuming a window size $M \times M \times M$, at each layer $l$ of the transformer encoder, the 3D tokens are evenly partitioned into $\left\lceil\frac{H}{2M}\right\rceil\times\left\lceil\frac{W}{2M}\right\rceil \times\left\lceil\frac{D}{2M}\right\rceil$ non-overlapping regions and the self-attention is computed within these windows. In the subsequent layer $l+1$, the partitioned windows are shifted by $(\frac{M}{2},\frac{M}{2},\frac{M}{2})$ tokens with respect to the self-attention in the previous layer. This window shifting mechanism allows for modeling the interaction between partitioned local regions. The output of a transformer block that consists of consecutive layers $l$ and $l+1$ is computed according to
\begin{equation}
\begin{array}{l}
\hat{{z}}^{l}=\text{W-MSA}(\text{LN}({z}^{l-1}))+{z}^{l-1} \\
{z}^{l}=\text{MLP}(\text{LN}(\hat{{z}}^{l}))+\hat{{z}}^{l} \\
\hat{{z}}^{l+1}=\text{SW-MSA}(\text{LN}({z}^{l}))+{z}^{l} \\
{z}^{l+1}=\text{MLP}(\text{LN}(\hat{{z}}^{l+1}))+\hat{{z}}^{l+1},
\end{array}
\label{eq:eq1}
\end{equation}
In which, $\text{SW-MSA}$ and $\text{W-MSA}$ denote shifted window and regular multi-head self-attention, $\text{LN}$ is the layer normalization and $\text{MLP}$ is a multilayer perceptron. Given a set of $Q, K, V$ denote queries, keys, and values, the self-attention within each local window is computed as in
\begin{equation}
    \textnormal{Attention}(Q, K, V) = \textnormal{Softmax}\left(\frac{QK^{\top}}{\sqrt{d}}+B\right)V,
\label{eq:eq2}
\end{equation}
Where $d$ denotes the feature size of the query and key and $B$ is the relative position bias.  

At the end of each stage, patch merging layers are used to downsample the resolution of feature representations. Furthermore. encoder and decoder of UNetFormer and UNetFormer+ are connected to each other via skip connections. In every resolution ${i}$ ($i \in \{0,1,2,3,4\})$ of the encoder, the output feature representations are reshaped into $\frac{H}{2^{i}} \times \frac{W}{2^{i}} \times \frac{D}{2^{i}}$ and sent to the decoder blocks. In UNetFormer+, the output of the encoder ($i=5$) is fed into the bottleneck consisting of two consecutive transformer layers as described in Eq.~\ref{eq:eq1}. The output of the bottleneck is then sent to the next decoder block. In every decoder block, features from the lower resolution are up-sampled (\textit{i.e.} trilinear interpolation) and concatenated with the skip connected features from the encoder. The final features ${w}^{i}$ in the decoder are processed in a layer projection according to 
\begin{equation}
\begin{array}{l}
{w}^{i}=\text{MLP}(\text{LN}({w}^{i}))\\
{w}^{i-1}=\text{W-MSA}(\text{LN}({w}^{i}))+{w}^{i}, \\
\end{array}
\label{eq:eq3}
\end{equation}
In UNetFormer, each convolutional block consists of two consecutive $3\times3\times3$ convolutional layers which are followed by instance normalization~\cite{ulyanov2016instance} and Leaky ReLU activation function. The bottleneck uses a convolutional block in lieu of two transformer layers. In the decoder block, a deconvolutional layer with kernel size of $2\times2\times2$ is used to increase the resolution by a factor of $2$. A convolutional block is used as the layer projection.

\subsection{Deep Supervision}
We utilize deep supervision for UNetFormer and UNetFormer+. Specifically, feature maps of resolution ${i}$ ($i \in \{1,2\})$ are up-sampled to the image resolution. The segmentation outputs, for all stages including the image resolution, are computed by using $1\times1\times1$ convolutional layers. Thus, the segmentation loss function $\mathcal{L}_{seg}$ is computed according to
\begin{equation}
\mathcal{L}_{seg} = \mathcal{L}(G_{0},Y_{0}) +\lambda_{1} \mathcal{L}(G_{1},Y_{1}) +\lambda_{2} \mathcal{L}(G_{2},Y_{2}),
\label{eq:loss_tot}
\end{equation}
$\lambda_{1}$ and $\lambda_{2}$ are set to $0.5$ and $0.25$ respectively. Each loss function is a combination of cross-entropy and soft Dice~\cite{milletari2016v} losses and computed as in 
\begin{equation}
\mathcal{L}(G,Y) = 1-\frac{2}{K}\sum_{k=1}^{K}\frac{\sum_{n=1}^{N} G_{n,k}Y_{n,k} }{\sum_{n=1}^{N}G^{2}_{n,k}+ \sum_{n=1}^{N}Y^{2}_{n,k}}
-\frac{1}{N}\sum_{n=1}^{N}\sum_{k=1}^{K} G_{n,k}\log Y_{n,k},
\label{eq:loss_func}
\end{equation}
In which $N$ and $K$ denote the number of voxels and semantic classes, respectively and $G_{n,k}$ and $Y_{n,k}$ are the probability outputs and their corresponding ground truth at voxel $n$ for class $k$.

\subsection{Pre-training Framework}
We present a framework which utilizes masked image modeling for self-supervised pre-training of the encoder in the proposed model. Specifically, assuming a given masking ratio and patch size, we randomly mask the 3D input volume and use it as an input to the encoder. In the proposed pre-training scheme, we utilize the encoder of our model, which is a 3D Swin Transformer and connect it to a lightweight CNN-based decoder via skip connections at multiple resolutions. Our design uses significantly less number of parameters in the decoder to speed up the pre-training and learn more effective visual representation via the encoder. Furthermore, the decoder reconstructs the masked tokens by leveraging the contextual information of visible tokens. We enforce a L1 mask reconstruction loss function $\mathcal{L}_{mask}$ by only considering the masked tokens according to  
\begin{equation}
\mathcal{L}_{mask}=\frac{1}{M}\left\|\mathbf{P}_{m}-\mathbf{X}_{m}\right\|_{1}.
\label{eq:eq4}
\end{equation}
Where $\mathbf{P}_{m}$ and $\mathbf{X}_{m}$ denote predictions of masked token and corresponding image tokens respectively and $M$ is the number of masked tokens. In addition, our proposed scheme which uses a skip-connected auto-encoder for pre-training allows for more accurate reconstructions which would have not been possible by simply upsampling to the original resolution and  using a linear layer as the reconstruction head.

\section{Experiments}
\subsection{Implementation Details}
We implemented our framework in PyTorch\footnote{\href{https://pytorch.org/}{https://pytorch.org/}} and MONAI\footnote{\href{https://monai.io/}{https://monai.io/}} and trained on a DGX-1 cluster with 4 NVIDIA V100 GPUs. The learning rate is set to $2\times10^{-4}$. The 3D Swin Transformer encoder has an embedding dimension of $C=96$. Each stage of the encoder and transformer-based decoder consists of $2$ and $1$ transformer layers respectively. We divide BraTS21~\cite{BratsAll2018,brats1,brats2,brats3,brats4} and MSD liver segmentation datasets~\cite{antonelli2021medical} by ratio of $75:20:5$ as for the training, testing and validation sets. Best model checkpoints on the validation set are used for inference. For MSD liver experiments, we use randomly cropped images with resolution $96 \times 96 \times 96$ and apply data augmentation transforms of random flip, rotation, intensity scaling and shifting. For BraTS 21 experiments, all input images are normalized to have zero mean and unit standard deviation by only considering the non-zero voxel values. For training, we use randomly cropped patches with resolution $128\times128\times128$ from the multi-modal 3D MRI images. For data augmentation, we employ random per channel intensity shift in the range $(-0.1,0.1)$, and random scaling of the intensity using values in the range of $(0.9, 1.1)$ for image input channels. In addition, random axis mirror flip by using a probability of $0.5$ for all major axes is utilized. The batch size per GPU was set to 1. All models were trained for a total of 500 epochs with a linear warmup and using a cosine annealing learning rate scheduler. Inference is conducted by utilizing a sliding window approach which has an overlapping of $0.7$ for neighboring voxels. 

\begin{table}[t]
\centering
	\caption{MSD dataset liver and liver tumor segmentation benchmarks using MSD dataset in terms of mean Dice score and Hausdorff Distance (HD). UNetFormer and UNetFormer+ use a CNN-based and transformer-based decoders respectively. Number of parameters are measured in millions. }
	\label{tab:liver_seg}
	\begin{tabular}{l|c|c|c|c|c|c|c}
		\hline
		& \multicolumn{2}{c|}{Dice} & \multicolumn{2}{c|}{HD}& \multicolumn{3}{c}{Considerations}  \\ \hline
		Models & Liver & Tumor & Liver & Tumor & \#Params& GFLOPs & Pre-trained \\ \hline
		V-Net~\cite{Milletari16} & 91.40 & 50.13 & 12.47 & 20.67 & 45.60& 161.92  & No\\
		nnFormer~\cite{zhou2021nnformer} & 92.07 & 50.02 & 11.86 & 21.52 & 149.04& 106.45  & Yes\\
		TransVW~\cite{haghighi2021transferable} & 93.27 & 54.17 & 11.13 & 13.84 & 19.06& 501.16  & Yes\\
		Models Genesis~\cite{zhou2021models} & 93.84 & 53.70 & 10.76 & 14.46 & 19.06& 501.16  & Yes\\
		SegResNet~\cite{Myronenko18} & 95.23 & 57.12 & 8.46 & 10.74 & \textbf{18.78} & 122.16  & No\\
		nnUNet~\cite{isensee2021nnu} & 95.67 & 57.97 & 7.94 & 9.28 & 33.07& 445.34  & No\\		
		UNetFormer+ & 94.79 & 50.32 & 9.93 & 18.98 & 24.44& \textbf{39.63}  & No\\
		UNetFormer+ & 95.06 & 51.23 & 9.18 & 17.41 & 24.44& \textbf{39.63}  & Yes\\
		UNetFormer & 95.73 & 58.05 & 7.68 & 9.13 & 58.96& 149.50  & No\\
		UNetFormer & \textbf{96.03} & \textbf{59.16} & \textbf{7.21} & \textbf{8.49} & 58.96& 149.50  & Yes  \\
		\hline
	\end{tabular}
\end{table}

\begin{table}[t]
\centering
	\caption{BraTS 21 dataset brain tumor segmentation benchmarks in terms of mean Dice score. ET, WT and TC denote Enhancing Tumor, Whole Tumor and Tumor Core respectively. UNetFormer and UNetFormer+ use a CNN-based and transformer-based decoders respectively. Number of parameters are measured in millions.}
	\label{tab:brats}
	\begin{tabular}{l|c|c|c|c|c|c}
		\hline
		& \multicolumn{4}{c|}{Dice} & \multicolumn{2}{c}{Considerations}  \\ \hline
		Models & ET & WT & TC & Avg. & \#Params & GFLOPs \\ \hline
		TransBTS~\cite{wang2021transbts} & 86.60& 90.30& 89.81 & 88.91 & 30.61 & 131.88 \\
		nnFormer~\cite{zhou2021nnformer} & 86.87& 92.68& 90.15 & 89.90 & 149.04 & 106.45 \\
		SegResNet~\cite{Myronenko18} & 88.40& 92.70& 91.70 & 90.90 & \textbf{18.78} & 122.16 \\
		nnUNet~\cite{isensee2021nnu} & 88.60& 92.91& 91.40 & 91.01 & 33.07 & 454.34 \\
		UNetFormer+ & 88.48& \textbf{93.67}& 91.89 & 91.20 & 24.44 & \textbf{39.63} \\
		UNetFormer & \textbf{88.80}& 93.22& \textbf{92.1} & \textbf{91.54} & 58.96 & 149.50 \\
		\hline
	\end{tabular}
\end{table}

\subsection{Results and Discussion}
\paragraph{Liver and Liver Tumor Segmentation}: We present the performance benchmarks of MSD liver and liver tumor segmentation task in Table~\ref{tab:liver_seg}. We have compared against both prominent CNN-based and transformer-based architectures such as nnUNet~\cite{isensee2021nnu} and nnFormer~\cite{zhou2021nnformer} in addition to pre-trained Models Genesis~\cite{zhou2021models} and TransVW~\cite{haghighi2021transferable} methodologies. Pre-trained UNetFormer outperforms nnUNet, which is the closest competing model, and nnFormer~\cite{zhou2021nnformer}, which is a 3D Swin Transformer-based model, by $4.30\%$ and $0.37\%$ for liver and $18.27\%$ and $3.57\%$ for liver tumor, respectively. In addition, randomly initialized UNetFormer outperforms both nnUNet and nnFormer in a similar manner. As a result, these benchmarks validate the effectiveness of our approach in segmenting and capturing the details of organs and tumors with various sizes.

\paragraph{Brain Tumor Segmentation}: We present the performance benchmarks of BraTS 21 brain tumor segmentation in Table~\ref{tab:brats}. We compared against previous state-of-the-art models on BraTS challenges with randomly initialized weights to demonstrate the generalizability of our proposed architecture across different tasks and imaging modalities. Both UNetFormer and UNetFormer+ outperform competing methodologies in terms of Dice score for all tumor regions. Specifically, UNetFormer outperforms nnUNet~\cite{isensee2021nnu}, nnFormer~\cite{zhou2021nnformer} and TransBTS~\cite{wang2021transbts}, which are CNN, Swin and ViT-based models, by $0.58\%$, $1.82\%$ and $2.95\%$ respectively, hence validating the effectiveness of our proposed model.  

\paragraph{Effectiveness of Pre-Training}: Our experiments show that pre-trained UNetFormer and UNetFormer+ models consistently outperform their randomly initialized counterparts for both liver and liver tumor segmentation, hence validating the effectiveness of representation learning in the pre-training framework. 

\paragraph{Performance of Decoder Designs}: Our experiments show that UNetFormer outperforms UNetFormer+ in most of the liver and brain tumor segmentation tasks. The UNetFormer+ is competitive in segmenting large-sized organs and tumors. Specifically, in brain tumor segmentation, it outperforms UNetFormer for segmenting the whole tumor regions. The superiority of UNetFormer for segmenting smaller organs could be attributed to the capability of its CNN layers in recovering localized information that may not be captured transformer layers. 

\paragraph{Efficiency and Accuracy}: In Tables~\ref{tab:liver_seg} and \ref{tab:brats}, we have also presented the computational requirements of our proposed models and other competing approaches. In terms of number of parameters, both UNetFormer and UNetFormer+ models are moderately sized when compared to larger models such as nnFormer. In addition, UNetFormer+ has the least GFLOPs in comparison to other approaches, hence allowing for higher computational efficiency with acceptable accuracy.

\begin{figure*}[t!]
\centering
\def\x{0.24}

\includegraphics[width=\x\linewidth,height=\x\linewidth]{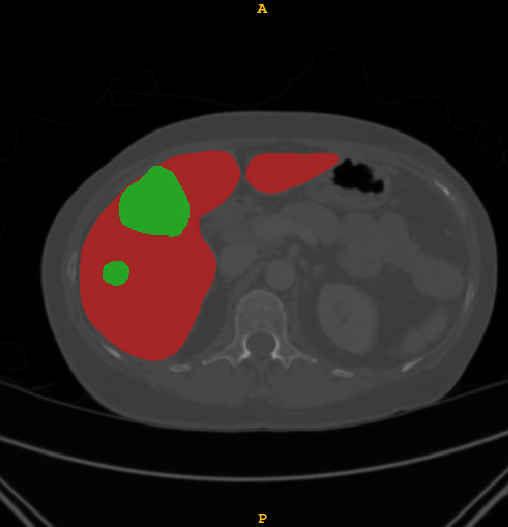}
\hfill
\includegraphics[width=\x\linewidth,height=\x\linewidth]{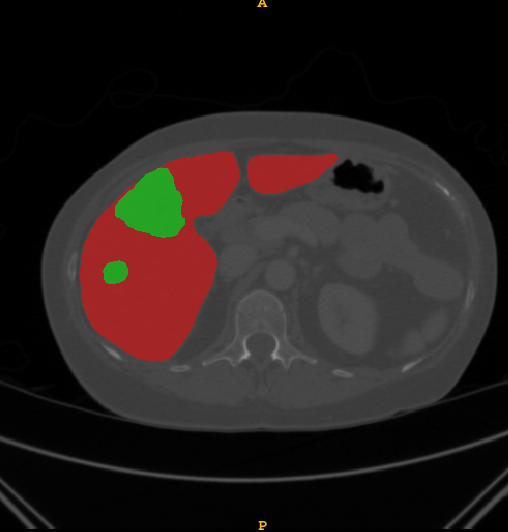}
\hfill
\includegraphics[width=\x\linewidth,height=\x\linewidth]{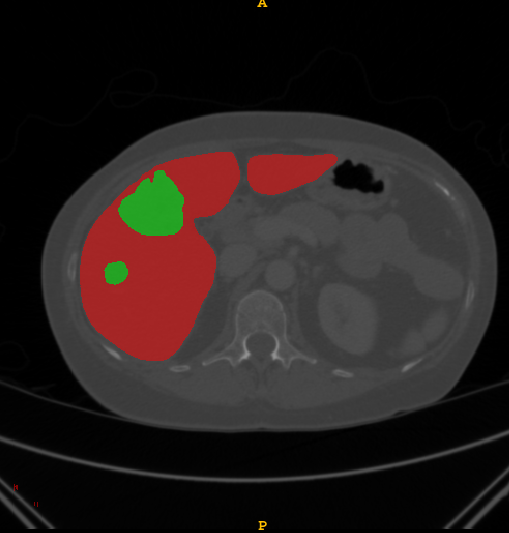}
\hfill
\includegraphics[width=\x\linewidth,height=\x\linewidth]{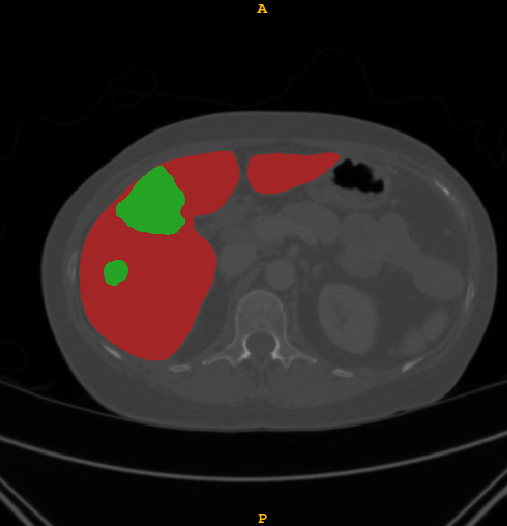}

\vspace{0.15pt}

\includegraphics[width=\x\linewidth,height=\x\linewidth]{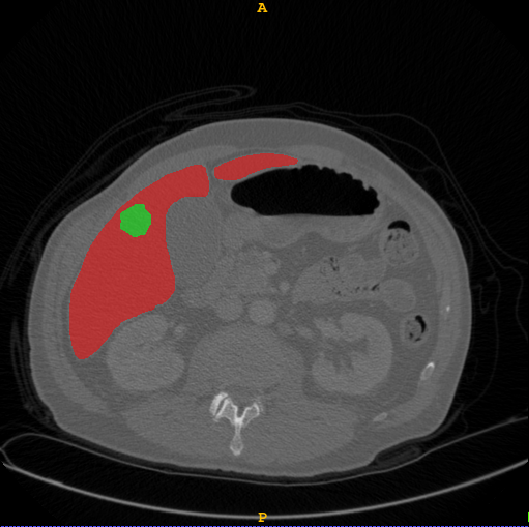}
\hfill
\includegraphics[width=\x\linewidth,height=\x\linewidth]{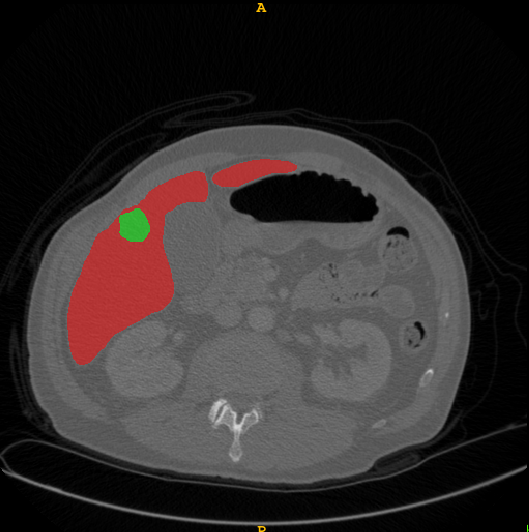}
\hfill
\includegraphics[width=\x\linewidth,height=\x\linewidth]{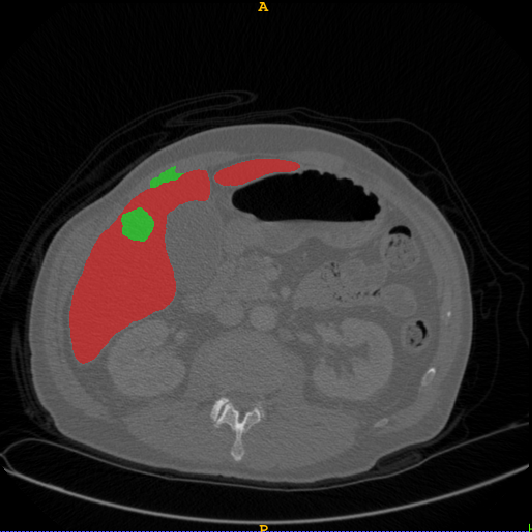}
\hfill
\includegraphics[width=\x\linewidth,height=\x\linewidth]{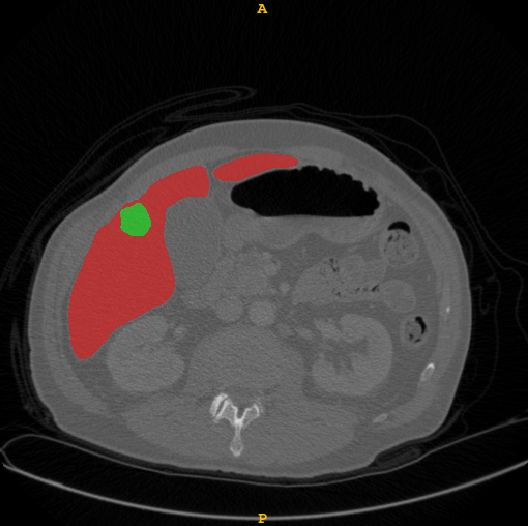}

\vspace{0.15pt}

\includegraphics[width=\x\linewidth,height=\x\linewidth]{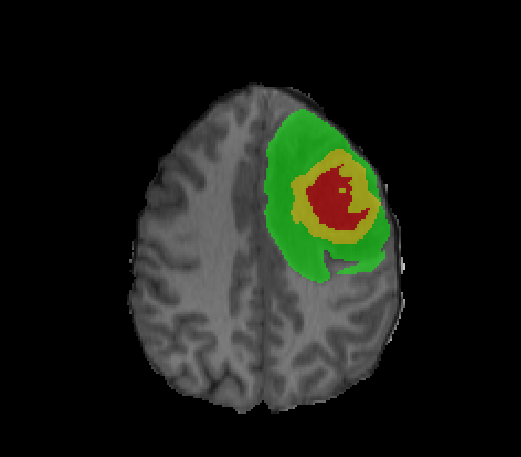}
\hfill
\includegraphics[width=\x\linewidth,height=\x\linewidth]{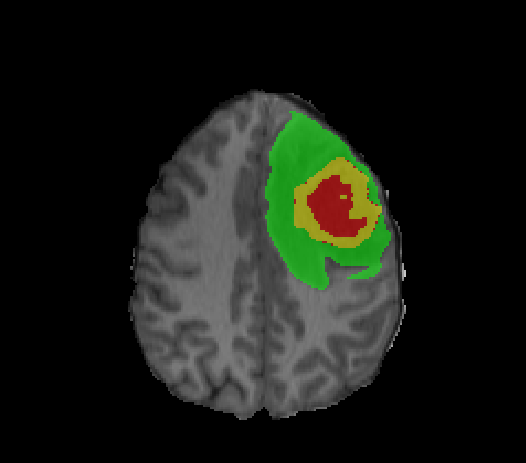}
\hfill
\includegraphics[width=\x\linewidth,height=\x\linewidth]{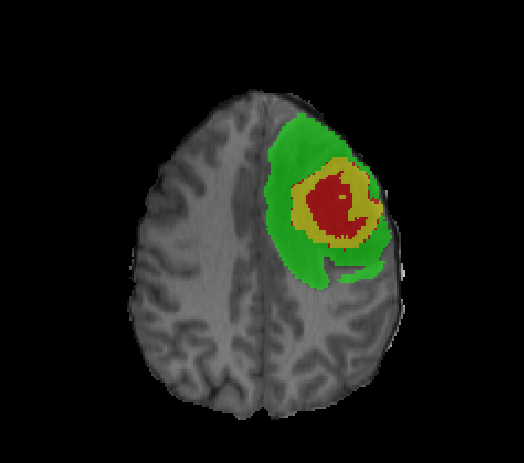}
\hfill
\includegraphics[width=\x\linewidth,height=\x\linewidth]{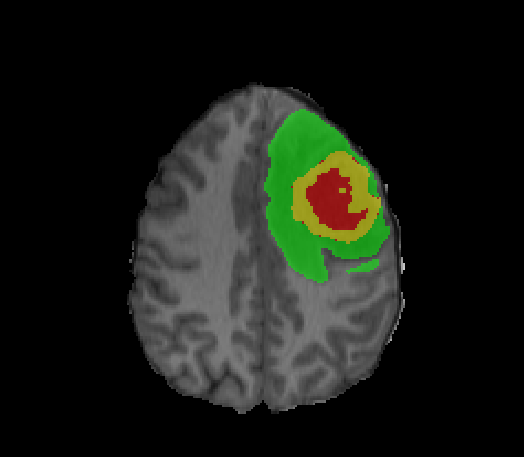}

\vspace{0.15pt}

\includegraphics[width=\x\linewidth,height=\x\linewidth]{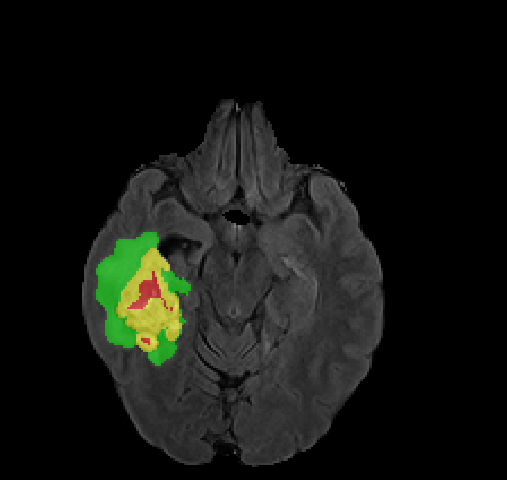}
\hfill
\includegraphics[width=\x\linewidth,height=\x\linewidth]{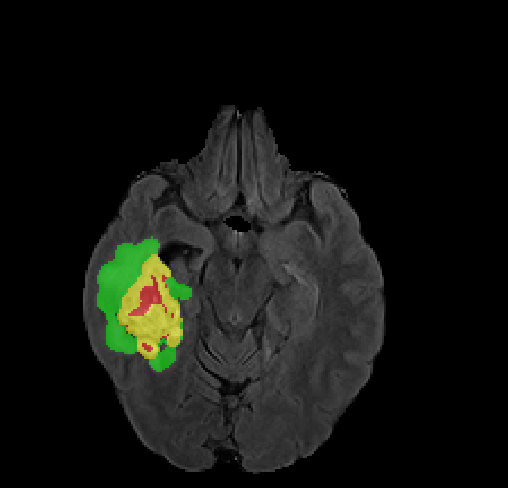}
\hfill
\includegraphics[width=\x\linewidth,height=\x\linewidth]{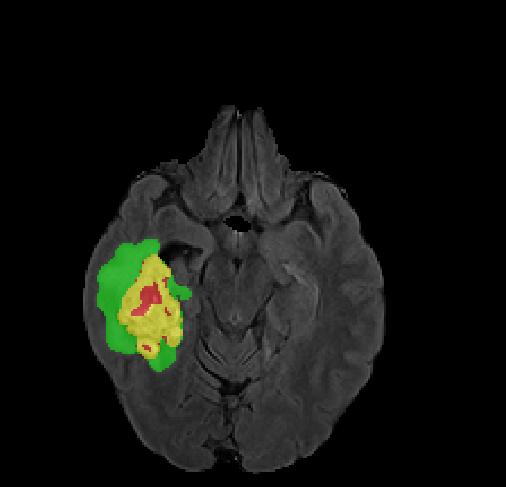}
\hfill
\includegraphics[width=\x\linewidth,height=\x\linewidth]{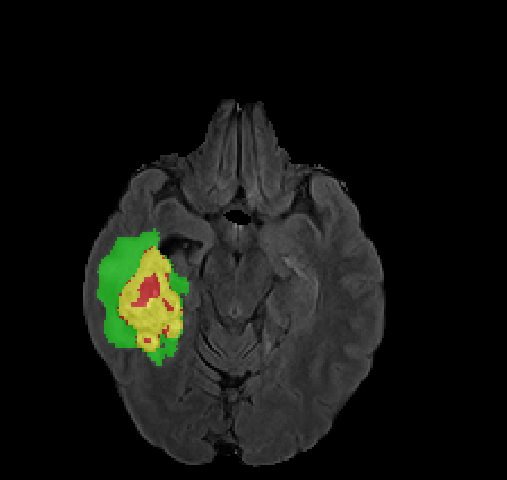}

\vspace{0.15pt}

\makebox[\x\linewidth]{(a)} \hfill \makebox[\x\linewidth]{(b)} \hfill
\makebox[\x\linewidth]{(c)} \hfill \makebox[\x\linewidth]{(d)}

\caption{(a) Ground Truth. Segmentation outputs of : (b) UNetFormer. (c) nnUNet. (d) SegResNet.}
\label{fig:attention_maps}
\end{figure*}

\begin{figure}[t]
\centering
\includegraphics[width=0.9\textwidth]{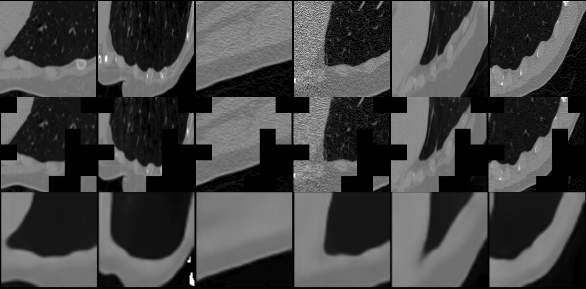}
  \caption{Visualization of pre-training reconstruction outputs. First, second and third rows from the top denote original, masked and reconstructed images, respectively.}
  \label{fig:recon}
\end{figure}

\subsection{Qualitative Segmentation Comparisons}
In Fig.~\ref{fig:attention_maps}, we present qualitative segmentation outputs of UNetFormer, nnUNet~\cite{isensee2021nnu} and SegResNet~\cite{Myronenko18} for the tasks of liver and liver tumor segmentation using MSD dataset~\cite{antonelli2021medical} as well as brain tumor segmentation using BraTS21 dataset~\cite{baid2021rsna}. As illustrated in these comparisons, the segmentation outputs of UNetFormer capture the fine-grained details of tumor structures, for both liver and brain segmentation tasks, in comparison to other models. In addition, segmentation outputs of other models suffer from artifacts.

\subsection{Pre-training Reconstruction Visualization}
We present the visualization of pre-training reconstruction outputs in addition to original and masked images in Fig.~\ref{fig:recon}. The reconstructed images are obtained for a masking ratio of $0.4$ and a patch size of $16$. These pre-training hyper-parameters are optimal in terms of down-stream segmentation performance in according to our analysis in Sec.~\ref{sec:abl}.

\begin{figure}[t]
\centering
\includegraphics[width=1\textwidth]{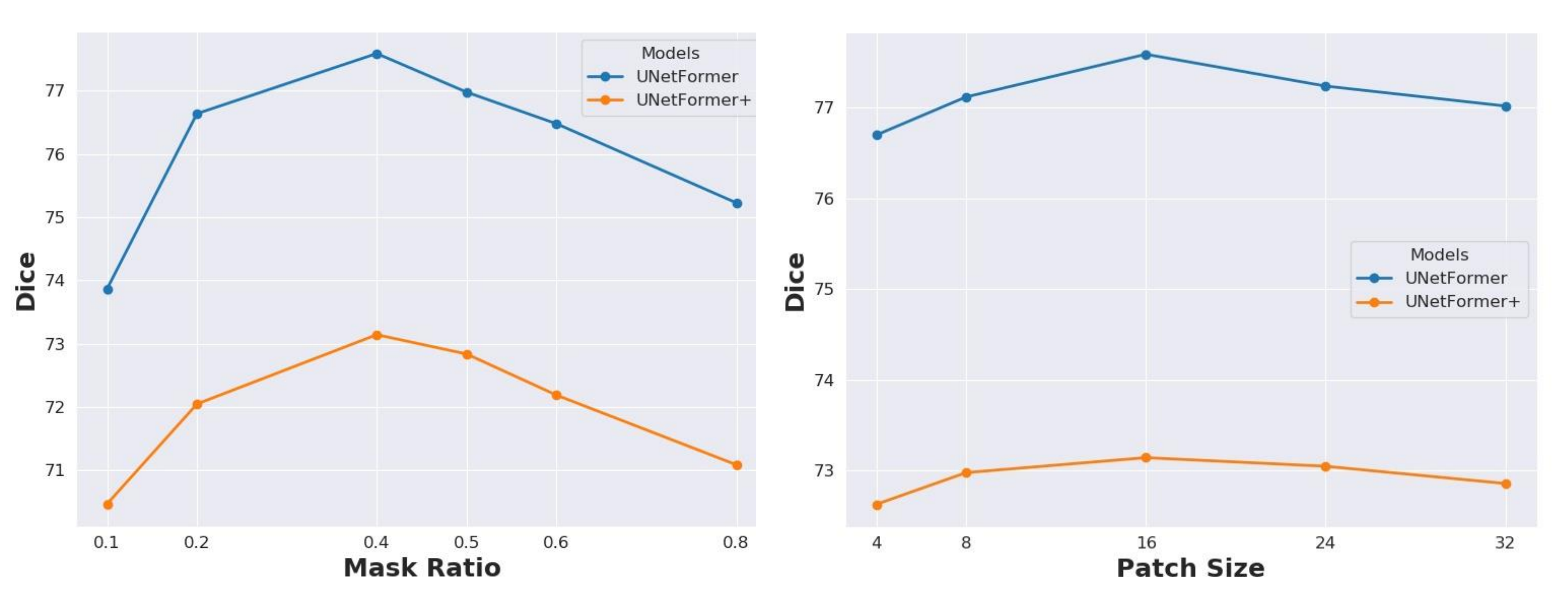}
  \caption{Effects of token masking ratio and patch size on the overall Dice accuracy. Dice values are computed as an average of liver and liver tumor segmentation performance on the MSD dataset.}
  \label{fig:pipeline2}
\end{figure}

\subsection{Ablation Studies}
\label{sec:abl}
In Fig.~\ref{fig:pipeline2}, we investigate the effects of token masking ratio and patch size on the average Dice score of liver and liver tumor segmentation performance. Moderate masking ratios, such as $40\%$, yields the best segmentation performance. We also study the impact of token patch size for effective self-supervised representation learning. Our analysis indicates that very large or small patch sizes are sub-optimal for token reconstruction and decreases the performance of downstream segmentation task. For an input volume size of $96 \times 96 \times 96$, our experiments show that a token patch size of $16 \times 16 \times 16$ results in the best downstream segmentation performance.

As shown in Fig.~\ref{fig:pipeline2}, the fine-tuning segmentation performance tends to be more sensitive to changes in the masking ratio as opposed to the patch size. In addition, the optimal patch size and masking ratio may need to be tuned for different domains and baseline encoder models. The correlation between image reconstruction quality and down stream fine-tuning tasks is outside the scope of the proposed effort and needs additional investigation.

\section{Conclusion}

In this paper, we introduced a novel unified vision transformer-based framework for volumetric medical image segmentation. Specifically, we proposed to utilize a 3D Swin Transformer-based encoder which is connected to a decoder via skip connections at multiple resolutions. We proposed UNetFormer and UNetFormer+ with CNN-based and transformer-based decoders respectively, that can effectively meet a variety of trade-off requirements between computational efficiency and model performance. We also proposed a self-supervised pre-training framework which utilizes masked image modeling as a proxy task. We validated the effectiveness of our proposed framework on liver and liver tumor segmentation task using the MSD dataset, and brain tumor segmentation task using BraTS 21 dataset. Our proposed framework achieves state-of-the-art performance on these tasks and by outperforming CNN, Swin and ViT-based segmentation baselines.

\bibliographystyle{splncs04}
\bibliography{ref}

\begin{thebibliography}{10}
\providecommand{\url}[1]{\texttt{#1}}
\providecommand{\urlprefix}{URL }
\providecommand{\doi}[1]{https://doi.org/#1}

\bibitem{antonelli2021medical}
Antonelli, M., Reinke, A., Bakas, S., Farahani, K., Landman, B.A., Litjens, G.,
  Menze, B., Ronneberger, O., Summers, R.M., van Ginneken, B., et~al.: The
  medical segmentation decathlon. arXiv preprint arXiv:2106.05735  (2021)

\bibitem{baid2021rsna}
Baid, U., Ghodasara, S., Mohan, S., Bilello, M., Calabrese, E., Colak, E.,
  Farahani, K., Kalpathy-Cramer, J., Kitamura, F.C., Pati, S., et~al.: The
  rsna-asnr-miccai brats 2021 benchmark on brain tumor segmentation and
  radiogenomic classification. arXiv preprint arXiv:2107.02314  (2021)

\bibitem{brats3}
Bakas, S., Akbari, H., Sotiras, A., Bilello, M., Rozycki, M., Kirby, J.,
  John~Freymann, K.F., Davatzikos, C.: Segmentation labels and radiomic
  features for the pre-operative scans of the tcga-gbm collection. The Cancer
  Imaging Archive  (2017), \url{https://doi.org/10.7937/K9/TCIA.2017.KLXWJJ1Q}

\bibitem{brats4}
Bakas, S., Akbari, H., Sotiras, A., Bilello, M., Rozycki, M., Kirby, J.,
  John~Freymann, K.F., Davatzikos, C.: Segmentation labels and radiomic
  features for the pre-operative scans of the tcga-lgg collection. The Cancer
  Imaging Archive  (2017), \url{https://doi.org/10.7937/K9/TCIA.2017.GJQ7R0EF}

\bibitem{brats2}
Bakas, S., Akbari, H., Sotiras, A., Bilello, M., Rozycki, M., Kirby, J.,
  Freymann, J., Farahani, K., Davatzikos, C.: Advancing the cancer genome atlas
  glioma {MRI} collections with expert segmentation labels and radiomic
  features. Scientific Data  \textbf{4} (2017)

\bibitem{BratsAll2018}
Bakas, S., Reyes, M., et~Int, Menze, B.: Identifying the best machine learning
  algorithms for brain tumor segmentation, progression assessment, and overall
  survival prediction in the {BRATS} challenge. In: arXiv:1811.02629 (2018)

\bibitem{bao2021beit}
Bao, H., Dong, L., Wei, F.: Beit: Bert pre-training of image transformers.
  arXiv preprint arXiv:2106.08254  (2021)

\bibitem{chen2021transunet}
Chen, J., Lu, Y., Yu, Q., Luo, X., Adeli, E., Wang, Y., Lu, L., Yuille, A.L.,
  Zhou, Y.: Transunet: Transformers make strong encoders for medical image
  segmentation. arXiv preprint arXiv:2102.04306  (2021)

\bibitem{cheng2021per}
Cheng, B., Schwing, A., Kirillov, A.: Per-pixel classification is not all you
  need for semantic segmentation. Advances in Neural Information Processing
  Systems  \textbf{34} (2021)

\bibitem{devlin2018bert}
Devlin, J., Chang, M.W., Lee, K., Toutanova, K.: Bert: Pre-training of deep
  bidirectional transformers for language understanding. arXiv preprint
  arXiv:1810.04805  (2018)

\bibitem{dosovitskiy2020image}
Dosovitskiy, A., Beyer, L., Kolesnikov, A., Weissenborn, D., Zhai, X.,
  Unterthiner, T., Dehghani, M., Minderer, M., Heigold, G., Gelly, S., et~al.:
  An image is worth 16x16 words: Transformers for image recognition at scale.
  In: International Conference on Learning Representations (2020)

\bibitem{haghighi2021transferable}
Haghighi, F., Taher, M.R.H., Zhou, Z., Gotway, M.B., Liang, J.: Transferable
  visual words: Exploiting the semantics of anatomical patterns for
  self-supervised learning. IEEE transactions on medical imaging  (2021)

\bibitem{hatamizadeh2022unetr}
Hatamizadeh, A., Tang, Y., Nath, V., Yang, D., Myronenko, A., Landman, B.,
  Roth, H.R., Xu, D.: Unetr: Transformers for 3d medical image segmentation.
  In: Proceedings of the IEEE/CVF Winter Conference on Applications of Computer
  Vision. pp. 574--584 (2022)

\bibitem{he2021masked}
He, K., Chen, X., Xie, S., Li, Y., Doll{\'a}r, P., Girshick, R.: Masked
  autoencoders are scalable vision learners. arXiv preprint arXiv:2111.06377
  (2021)

\bibitem{isensee2021nnu}
Isensee, F., Jaeger, P.F., Kohl, S.A., Petersen, J., Maier-Hein, K.H.: nnu-net:
  a self-configuring method for deep learning-based biomedical image
  segmentation. Nature Methods  \textbf{18}(2),  203--211 (2021)

\bibitem{liu2021swin}
Liu, Z., Lin, Y., Cao, Y., Hu, H., Wei, Y., Zhang, Z., Lin, S., Guo, B.: Swin
  transformer: Hierarchical vision transformer using shifted windows. In:
  Proceedings of the IEEE/CVF International Conference on Computer Vision. pp.
  10012--10022 (2021)

\bibitem{brats1}
Menze, B.H., Jakab, A., Bauer, S., Kalpathy-Cramer, J., Farahani, K., Kirby,
  J., Burren, Y., Porz, N., Slotboom, J., Wiest, R., Lanczi, L., Gerstner,
  E.R., Weber, M.A., Arbel, T., Avants, B.B., Ayache, N., Buendia, P., Collins,
  D.L., Cordier, N., Corso, J.J., Criminisi, A., Das, T., Delingette, H.,
  Demiralp, C., Durst, C.R., Dojat, M., Doyle, S., Festa, J., Forbes, F.,
  Geremia, E., Glocker, B., Golland, P., Guo, X., Hamamci, A., Iftekharuddin,
  K.M., Jena, R., John, N.M., Konukoglu, E., Lashkari, D., Mariz, J.A., Meier,
  R., Pereira, S., Precup, D., Price, S.J., Raviv, T.R., Reza, S.M.S., Ryan,
  M.T., Sarikaya, D., Schwartz, L.H., Shin, H.C., Shotton, J., Silva, C.A.,
  Sousa, N., Subbanna, N.K., Szekely, G., Taylor, T.J., Thomas, O.M., Tustison,
  N.J., Unal, G.B., Vasseur, F., Wintermark, M., Ye, D.H., Zhao, L., Zhao, B.,
  Zikic, D., Prastawa, M., Reyes, M., Leemput, K.V.: The multimodal brain tumor
  image segmentation benchmark (brats). IEEE Trans. Med. Imaging
  \textbf{34}(10),  1993--2024 (2015)

\bibitem{milletari2016v}
Milletari, F., Navab, N., Ahmadi, S.A.: V-net: Fully convolutional neural
  networks for volumetric medical image segmentation. In: 2016 fourth
  international conference on 3D vision (3DV) (2016)

\bibitem{Milletari16}
Milletari, F., Navab, N., Ahmadi, S.A.: V-net: Fully convolutional neural
  networks for volumetric medical image segmentation. In: Fourth International
  Conference on 3D Vision (3DV) (2016)

\bibitem{Myronenko18}
Myronenko, A.: {3D} {MRI} brain tumor segmentation using autoencoder
  regularization. In: {BrainLes}, Medical Image Computing and Computer Assisted
  Intervention {(MICCAI)}. pp. 311--320. LNCS, Springer (2018)

\bibitem{raghu2021vision}
Raghu, M., Unterthiner, T., Kornblith, S., Zhang, C., Dosovitskiy, A.: Do
  vision transformers see like convolutional neural networks? Advances in
  Neural Information Processing Systems  \textbf{34} (2021)

\bibitem{shamshad2022transformers}
Shamshad, F., Khan, S., Zamir, S.W., Khan, M.H., Hayat, M., Khan, F.S., Fu, H.:
  Transformers in medical imaging: A survey. arXiv preprint arXiv:2201.09873
  (2022)

\bibitem{ulyanov2016instance}
Ulyanov, D., Vedaldi, A., Lempitsky, V.: Instance normalization: The missing
  ingredient for fast stylization. arXiv preprint arXiv:1607.08022  (2016)

\bibitem{wang2021transbts}
Wang, W., Chen, C., Ding, M., Yu, H., Zha, S., Li, J.: Transbts: Multimodal
  brain tumor segmentation using transformer. In: International Conference on
  Medical Image Computing and Computer-Assisted Intervention. pp. 109--119.
  Springer (2021)

\bibitem{xie2021simmim}
Xie, Z., Zhang, Z., Cao, Y., Lin, Y., Bao, J., Yao, Z., Dai, Q., Hu, H.:
  Simmim: A simple framework for masked image modeling. arXiv preprint
  arXiv:2111.09886  (2021)

\bibitem{zhou2021nnformer}
Zhou, H.Y., Guo, J., Zhang, Y., Yu, L., Wang, L., Yu, Y.: nnformer: Interleaved
  transformer for volumetric segmentation. arXiv preprint arXiv:2109.03201
  (2021)

\bibitem{zhou2021models}
Zhou, Z., Sodha, V., Pang, J., Gotway, M.B., Liang, J.: Models genesis. Medical
  image analysis  \textbf{67},  101840 (2021)

\end{thebibliography}

\end{document}


\title{UNetFormer: A Unified Vision Transformer Model and Pre-Training Framework for 3D Medical Image Segmentation \\ 
Supplementary Material}
\titlerunning{Supplementary Material}
%
%
\author{Anonymous}
\authorrunning{Anonymous et al.}
\maketitle              

\begin{figure}[H]
\centering
\includegraphics[width=0.9\textwidth]{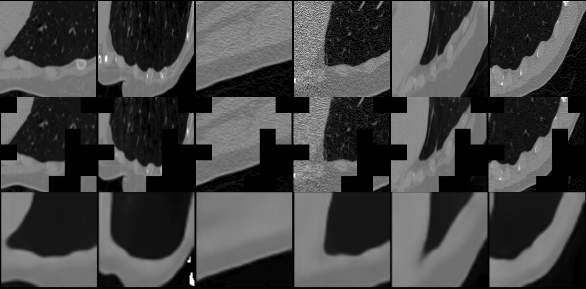}
  \caption{Pre-training reconstruction outputs. First, second and third rows denote original, masked and recontructed images, respectively.}
  \label{fig:recon}
\end{figure}

\begin{figure*}[t!]
\def\x{0.24}

\includegraphics[width=\x\linewidth,height=\x\linewidth]{Images/supplementary/case1_gt.png}
\hfill
\includegraphics[width=\x\linewidth,height=\x\linewidth]{Images/supplementary/case1_unetformer.png}
\hfill
\includegraphics[width=\x\linewidth,height=\x\linewidth]{Images/supplementary/case1_nnunet.png}
\hfill
\includegraphics[width=\x\linewidth,height=\x\linewidth]{Images/supplementary/case1_segresnet.png}

\vspace{0.15pt}

\includegraphics[width=\x\linewidth,height=\x\linewidth]{Images/supplementary/gt_case128.png}
\hfill
\includegraphics[width=\x\linewidth,height=\x\linewidth]{Images/supplementary/unetformer_case128.png}
\hfill
\includegraphics[width=\x\linewidth,height=\x\linewidth]{Images/supplementary/nnunet_case128.png}
\hfill
\includegraphics[width=\x\linewidth,height=\x\linewidth]{Images/supplementary/segresnet_case_128.png}

\vspace{0.15pt}

\includegraphics[width=\x\linewidth,height=\x\linewidth]{Images/supplementary/case_2_brats_gt.png}
\hfill
\includegraphics[width=\x\linewidth,height=\x\linewidth]{Images/supplementary/case2_brats_unetformer.png}
\hfill
\includegraphics[width=\x\linewidth,height=\x\linewidth]{Images/supplementary/case2_brats_nnunet.png}
\hfill
\includegraphics[width=\x\linewidth,height=\x\linewidth]{Images/supplementary/case2_brats_segresnet.png}

\vspace{0.15pt}

\includegraphics[width=\x\linewidth,height=\x\linewidth]{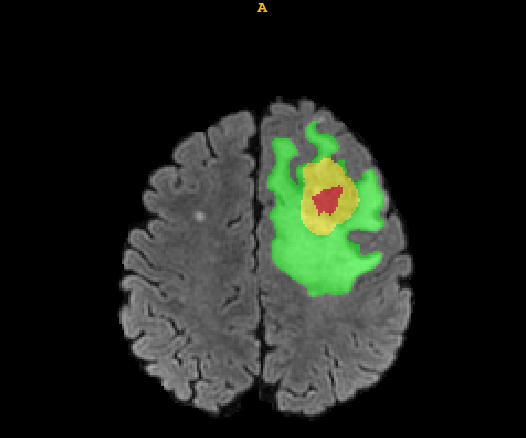}
\hfill
\includegraphics[width=\x\linewidth,height=\x\linewidth]{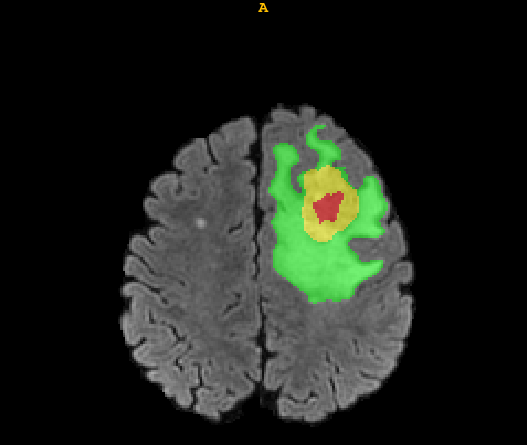}
\hfill
\includegraphics[width=\x\linewidth,height=\x\linewidth]{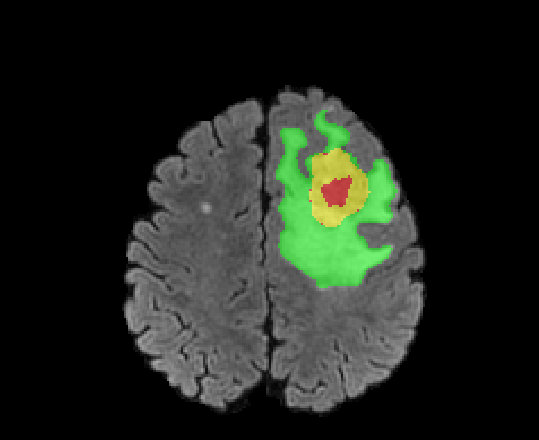}
\hfill
\includegraphics[width=\x\linewidth,height=\x\linewidth]{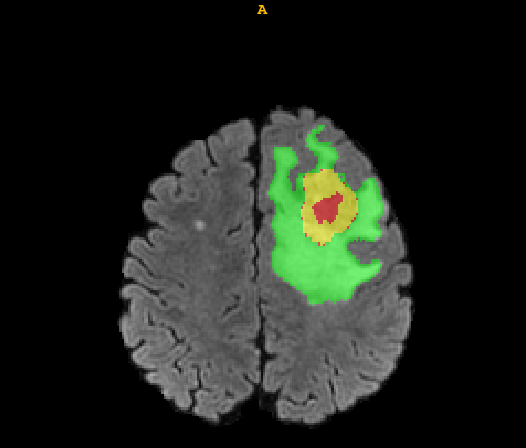}

\vspace{0.15pt}

\includegraphics[width=\x\linewidth,height=\x\linewidth]{Images/supplementary/gt_3_brats.png}
\hfill
\includegraphics[width=\x\linewidth,height=\x\linewidth]{Images/supplementary/unetformer_3_brats.png}
\hfill
\includegraphics[width=\x\linewidth,height=\x\linewidth]{Images/supplementary/nnunet_3_brats.png}
\hfill
\includegraphics[width=\x\linewidth,height=\x\linewidth]{Images/supplementary/segresnet3_brats.png}

\vspace{0.15pt}

\makebox[\x\linewidth]{(a)} \hfill \makebox[\x\linewidth]{(b)} \hfill
\makebox[\x\linewidth]{(c)} \hfill \makebox[\x\linewidth]{(d)}

\caption{(a) Ground Truth. Segmentation outputs of : (b) UNetFormer. (c) nnUNet. (d) SegResNet.}
\label{fig:attention_maps}
\end{figure*}

\section{Reconstruction Outputs}
We present pre-training reconstruction outputs in addition to original and masked images in Fig.~\ref{fig:recon}. The reconstructed images are obtained for a masking ratio of $0.4$ and a patch size of $16$, which have shown to be the most optimal pre-training setting in terms of down-stream segmentation performance in our ablation studies.

\section{Details of Pre-training Framework}
In our pre-training setup, we utilize the encoder of UNetFormer, which is a 3D Swin Transformer and connect it to a lightweight CNN-based decoder via skip connections at multiple resolutions. The decoder resembles the proposed CNN-based decoder but with significantly less number of parameters to speed up the pre-training process and learn more effective visual representation via the encoder. We also note that our proposed scheme which uses a skip-connected auto-encoder for pre-training allows for more accurate reconstructions which would have not been possible by simply upsampling to the original resolution and  using a linear layer as the reconstruction head.

\section{Qualitative Segmentation Outputs}
In Fig.~\ref{fig:attention_maps}, we present qualitative segmentation outputs of UNetFormer, nnUNet~\cite{isensee2021nnu} and SegResNet~\cite{Myronenko18} for the tasks of liver and liver tumor segmentation using MSD dataset~\cite{antonelli2021medical} as well as brain tumor segmentation using BraTS21 dataset~\cite{baid2021rsna}. As illustrated in these comparisons, UNetFormer segmentation outputs capture the fine-grained details of the tumor structures, for both liver and brain segmentation tasks, in comparison to other models. In addition, segmentation outputs of other models suffer from artifacts.

\bibliographystyle{splncs04}
\bibliography{ref}